\documentclass[a4paper,fleqn,usenatbib]{mnras}

\usepackage{newtxtext,newtxmath}
\usepackage{longtable}

\usepackage[T1]{fontenc}
\usepackage{ae,aecompl}

\usepackage{graphicx}	
\usepackage{amsmath}	
\usepackage{amssymb}	
\usepackage{ulem}
\usepackage{float}
\usepackage{placeins}

\title[VVV-WIT-07]{VVV-WIT-07: another Boyajian's
  star  or  a  Mamajek's object?\thanks{Based  on  observations  taken
    within  the  ESO  Public  Surveys  VVV  and  VVVX,  Programme  IDs
    179.B-2002  and  198.B-2004,  respectively,  and  on  observations
    obtained at the Southern  Astrophysical Research (SOAR) telescope,
    which is a joint project  of the Min.  da Ci\^{e}ncia, Tecnologia,
    Inova\c{c}\~{o}es  e Comunica\c{c}\~{o}es  (MCTIC) do  Brasil, the
    U.S.    National  Optical   Astronomy   Observatory  (NOAO),   the
    University of  North Carolina at  Chapel Hill (UNC),  and Michigan
    State University (MSU).}}

\author[R.~K.~Saito et al.]{R.~K.~Saito$^{1}$,\thanks{E-mail: saito@astro.ufsc.br}
D.~Minniti$^{2,3,4}$,
V.~D.~Ivanov$^{5}$,
M.~Catelan$^{6,3}$\thanks{On  sabbatical leave  at the  Astronomisches
  Rechen-Institut,  Zentrum  f\"{u}r  Astronomie  der  Universit\"{a}t
  Heidelberg, M\"{o}nchhofstr.  12-14,  69120 Heidelberg, Germany, and
  European  Southern  Observatory,  Av.   Alonso  de  C\'ordova  3107,
  7630355 Vitacura, Santiago, Chile.},
F.~Gran$^{6,3}$,
\newauthor
R.~Baptista$^{1}$,
R.~Angeloni$^{7,8}$,
C.~Caceres$^{2,9}$,
J.~C.~Beamin$^{10,11}$
\\
\\
$^{1}$Departamento de  F\'{i}sica, Universidade  Federal de
  Santa Catarina, Trindade 88040-900, Florian\'opolis, SC, Brazil\\
$^{2}$Departamento  de  Fisica,  Facultad  de  Ciencias  Exactas,
  Universidad  Andres Bello,  Av.  Fernandez  Concha 700,  Las Condes,
  Santiago, Chile\\
$^{3}$Instituto Milenio de Astrof\'isica, Santiago, Chile\\
$^{4}$Vatican Observatory, V00120 Vatican City State, Italy\\
$^{5}$European   Southern   Observatory,  Karl-Schwarszchild-Str.   2,
  D-85748 Garching bei Muenchen, Germany\\
$^{6}$Instituto  de Astrof\'isica,  Pontificia  Universidad Cat\'olica  de
  Chile, Av. Vicu\~na Mackenna 4860, 782-0436 Macul, Santiago, Chile\\
$^{7}$Instituto  de  Investigaci\'on  Multidisciplinar  en  Ciencia  y
  Tecnolog\'ia,  Universidad de  La  Serena,  Avenida Ra\'ul  Bitr\'an
  1305, La Serena, Chile\\
$^{8}$Departamento de  F\'isica y Astronom\'ia, Universidad  de La Serena,
  Avenida Cisternas 1200 Norte, La Serena, Chile\\
$^{9}$ N\'ucleo  Milenio Formaci\'on Planetaria -  NPF, Universidad de
  Valpara\'iso, Av. Gran Breta\~na 1111, Valpara\'iso, Chile\\
$^{10}$Instituto  de F\'isica  y Astronom\'ia,  Facultad de  Ciencias,
  Universidad de Valpara\'iso, Ave.  Gran Breta\~na 1111, Playa Ancha,
  Valpara\'iso, Chile.\\
$^{11}$N\'ucleo    Astroqu\'imica   y    Astrof\'isica,   Facultad    de
  Ingenier\'ia, Universidad Aut\'onoma de Chile, Chile
}

\date{Accepted XXX. Received YYY; in original form ZZZ}

\pubyear{2015}

\begin{document}
\label{firstpage}
\pagerange{\pageref{firstpage}--\pageref{lastpage}}
\maketitle

\begin{abstract}

We  report  the discovery  of  VVV-WIT-07,  an unique  and  intriguing
variable source presenting a sequence  of recurrent dips with a likely
deep eclipse  in July 2012.   The object was found  serendipitously in
the near-IR data obtained by the VISTA Variables in the V\'ia L\'actea
(VVV) ESO  Public Survey.  Our  analysis is based on  VVV variability,
multicolor, and proper motion (PM)  data.  Complementary data from the
VVV eXtended survey  (VVVX) as well as archive  data and spectroscopic
follow-up  observations aided  in the  analysis and  interpretation of
VVV-WIT-07.  A search  for  periodicity in  the  VVV $K_{\rm  s}$-band
light curve of VVV-WIT-07 results in  two tentative periods at $P\sim$
322 days  and $P\sim$ 170  days. Colors  and PM are  consistent either
with a reddened MS  star or a pre-MS star in  the foreground disk. The
near-IR spectra of VVV-WIT-07  appear featureless, having no prominent
lines in emission or absorption.  Features found in the light curve of
VVV-WIT-07 are  similar to those  seen in J1407 (Mamajek's  object), a
pre-MS  K5   dwarf  with  a   ring  system  eclipsing  the   star  or,
alternatively,  to KIC  8462852  (Boyajian's star),  an  F3 IV/V  star
showing irregular  and aperiodic dips in  its light curve. Alternative
scenarios, none of which is  fully consistent with the available data,
are  also briefly  discussed, including  a young  stellar object,  a T
Tauri star surrounded  by clumpy dust structure, a  main sequence star
eclipsed by a nearby extended  object, a self-eclipsing R CrB variable
star, and even a long-period, high-inclination X-ray binary.

\end{abstract}




\section{Introduction}
\label{sec:intro}

In recent  years, a number of  variable stars have been  identified in
ongoing wide-field variability  surveys, which have given  rise to new
classes  of variability  and  expanded our  views  about the  variable
Universe.   Examples of  that are  the Slowly  Pulsating B-type  Stars
\citep[SPBs;][]{2013A&A...554A.108M},    the   Blue    Large-Amplitude
Pulsators   \citep[BLAPs;][]{2017MNRAS.465.3039C}  and   the  McNeil's
Nebular Objects \citep[MNors;][]{2017NatAs...1E.166P}.   Some of these
objects  are  rare/unique  in  their   nature,  such  as  KIC  8462852
\citep[the Boyajian's star,][]{2016MNRAS.457.3988B} and J1407 \citep[=
  V1400  Cen, ][]{2012AJ....143...72M}.   Synoptic  surveys are  major
contributors to  these discoveries and  much more will  certainly come
out with the  advent of LSST \citep{2008arXiv0805.2366I}  and the next
generation   of    time-series   space   missions,    including   TESS
\citep{2015JATIS...1a4003R} and PLATO \citep{2014ExA....38..249R}.

The VISTA Variables in the V\'ia  L\'actea (VVV) is an ESO variability
survey of the inner  Milky Way, which mapped about 562  sq. deg in the
bulge           and          southern           Galactic          disk
\citep{2010NewA...15..433M,2012A&A...537A.107S}. Focused  on unveiling
the  3-D structure  of  the Milky  Way using  pulsating  RR Lyrae  and
Cepheid variables as distance indicators,  the VVV data are also being
mined on the  search for microlensing events,  eclipsing binaries, pre
main sequence (MS) variables, etc.   In 2016 a complementary survey to
VVV  called VVV  eXtended Survey  \citep[VVVX,][]{2018ASSP...51...63M}
started observations, including revisiting  the original VVV area thus
expanding the  original time  baseline and increasing  the photometric
depth,  in   addition  to   affording  improved  proper   motion  (PM)
measurements, as a result of combining both the VVV and VVVX datasets.

Among the targets found as variable  sources in the VVV data, some are
specially important  since their  behaviour does not  seem to  fit any
currently  known  class of  stellar  variability.   These objects  are
labeled as  ``What Is This''  (WIT) objects.  Most of  VVV-WIT objects
found up to now are high amplitude variables, including a transient of
unknown character, proposed to be  a nearby supernova, a rare Galactic
nova, or a  stellar merger \citep[VVV-WIT-06,][]{2017ApJ...849L..23M}.
Here we present the case  of VVV-WIT-07, an intriguing variable source
and unique in its nature found in the current VVV data.

\begin{figure}
\centering
\includegraphics[scale=0.35]{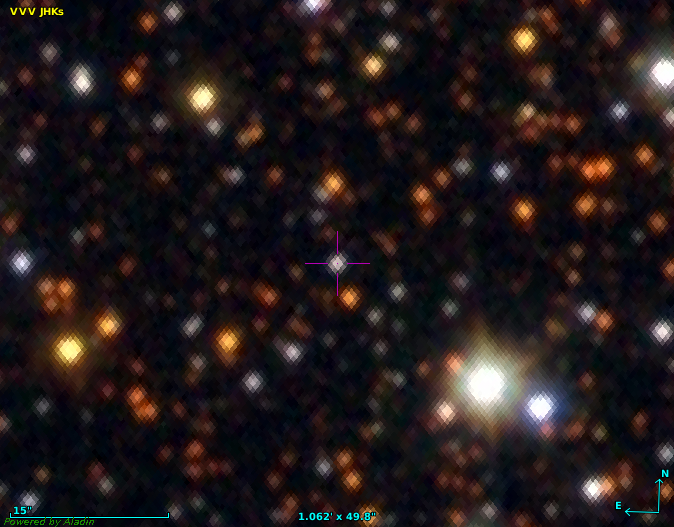}
\caption{VVV $JHK_{\rm s}$ false-color image of VVV-WIT-007 area.  The
  field  size  is  $60''  \times  50''$  and  oriented  in  equatorial
  coordinates. North  is towards  the top and  East towards  the left.
  The reticle at the centre marks VVV-WIT-007.}
\label{fig:image}
\end{figure}

\section{Observations and archive data}
\label{sec:obs}

\begin{figure*}
\centering
\includegraphics[scale=1.6]{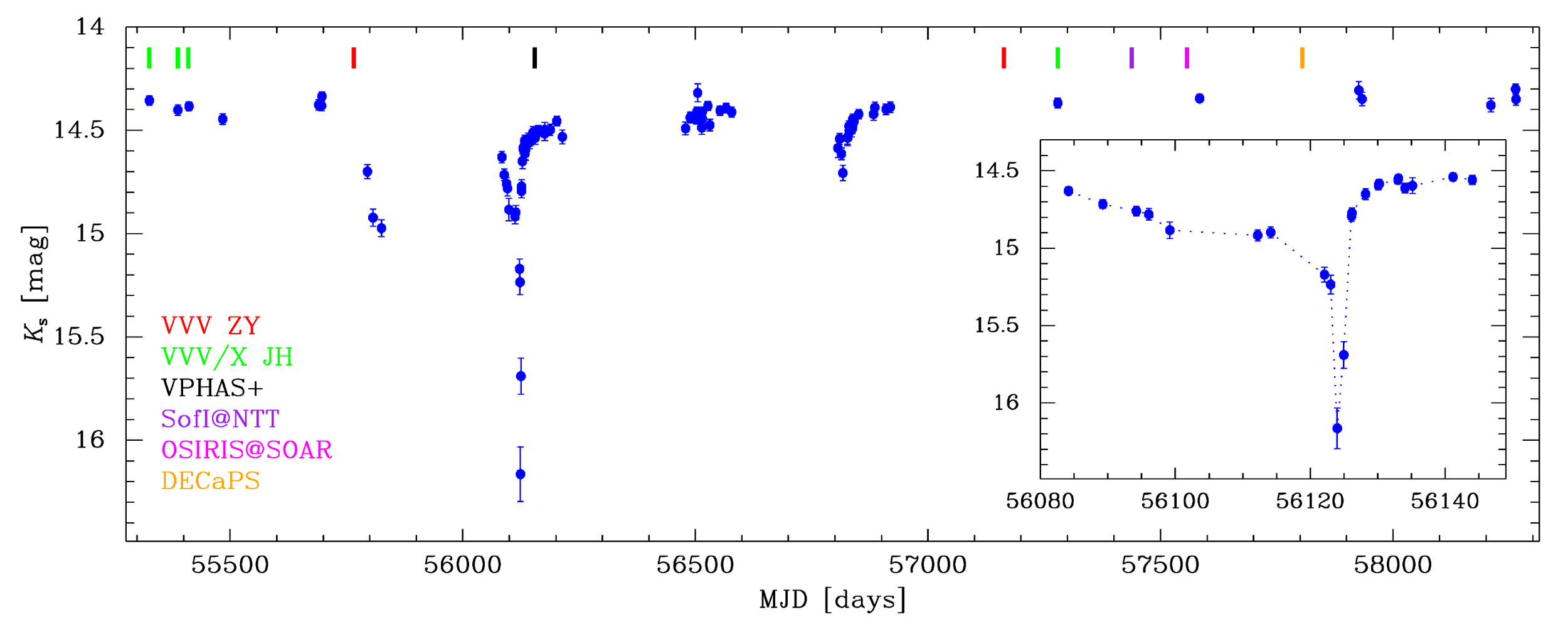}
\caption{VVV $K_{\rm s}$-band light curve  of VVV-WIT-007.  There is a
  total of  85 data-points  covering the 2010-2018  seasons, including
  data from  the VVVX  survey. Epochs of  contemporary multiwavelength
  observations as  well as  of spectroscopic observations  are marked.
  The insert shows an expanded view around the event in July 2012.}
\label{fig:lcurve}
\end{figure*}

VVV data consist  of two sets of casi-simultaneous  $ZY$ and $JHK_{\rm
  s}$ photometry, and  a variability campaign in  the $K_{\rm s}$-band
with $50-200$  epochs carried out  over many years  ($2010-2016$). The
strategy of  the VVVX  Survey is  similar to the  VVV and  consists of
$JHK_{\rm s}$ photometry plus 3 to  10 epochs in $K_{\rm s}$-band.  In
particular for  the field where  VVV-WIT-07 is located,  and combining
VVV and VVVX data,  2 sets of $ZY$ and 4 sets  of $JHK_{\rm s}$ images
have  been observed.   In addition,  85 $K_{\rm  s}$-band epochs  were
taken from May 10 2010 to May 25 2018 with irregular cadence.  The VVV
and  VVVX data  presented  here  are based  on  the default  ``aper3''
photometry provided  by the Cambridge Astronomical  Survey Unit (CASU)
on    the    stacked    VVV    tile    images    \citep[for    details
  see][]{2012A&A...537A.107S}. The  set of 85 $K_{\rm  s}$-band epochs
from VVV and  VVV-X available for VVV-WIT-07 is  presented in Appendix
A.

VVV-WIT-07  is a  stellar  source  located in  the  Galactic plane  at
coordinates RA, DEC (J2000)=17:26:29.387,$-$35:40:56.20, corresponding
to  $l,b$=$-$7.8580,$-$0.2357~deg.  Figure~\ref{fig:image}  shows that
the object is a relatively  isolated point source (i.e.  not blended),
and seems to be bluer than  the surrounding fainter field stars, which
appear  to be  more  affected  by reddening.   This  object was  found
serendipitously in  a search  for large amplitude  objects in  the VVV
data (e.g.  novae and LPVs). It stood out in our search because it had
a large amplitude dip in  2012.  The VVV-WIT-07 $K_{\rm s}$-band light
curve  presents  an  irregular  behaviour with  a  main  deep,  narrow
eclipse/dip of $\Delta K_{\rm s}  \sim1.8$~mag delayed with respect to
a    broad    and    shallower    dip   around    July    2012    (see
Fig.~\ref{fig:lcurve}). The narrow  eclipse/dip lasts by $\sim11$~days
while  the broad  dip  is seen  with a  width  of $\sim48$~days.   The
variation  in  magnitude  from  the median  <$K_{\rm  s}$>=  14.35~mag
measured over the  other epochs implies that $\sim$80$\%$  of the flux
in $K_{\rm s}$ is missing during the event on July 2012.

Archive search  at the VVV-WIT-07 position  shows several measurements
in  different  wavelengths spanning  from  Gaia  data in  the  optical
\citep{2018A&A...616A...1G}  to  GLIMPSE  observations in  the  mid-IR
\citep{2009yCat.2293....0S}          .           Gaia,          DECaPS
\citep{2017arXiv171001309S}  and   VPHAS+  \citep{2016yCat.2341....0D}
observations  are  contemporaneous  with our  VVV/VVVX  data,  however
epochs    for    Gaia    observations    are    not    yet    publicly
available\footnote{Individual  epoch data  for  the Gaia  observations
  will  be released  only with  the final  catalogue, as  described in
  cosmos.esa.int/web/gaia/release}.   The multi-wavelength  dataset is
presented  in  Table   1  and  a  photospectrum   combining  all  this
information is shown in the top panel of Fig.~\ref{fig:sed}.

Given the interesting unknown nature of this object, follow-up near-IR
spectra of VVV-WIT-07 were taken with the SofI spectrograph at ESO NTT
telescope on  UT February 02  2016 (Grism Red  - GR, $J$  to $K$-band,
$\lambda \sim  1.5 -2.5~ \mu m$)  and with the OSIRIS  spectrograph at
the SOAR Telescope on UT June 18 2016 ($K$-band, $R\sim1200$, $\lambda
\sim  2.0 -2.3~  \mu  m$).  Both  data sets  were  data reduced  using
standard IRAF tasks.  The resulting spectra are basically flat, with a
continuum  without  prominent lines  in  emission  or absorption  (see
Fig.~\ref{fig:sed}). The  only likely  features seem to  correspond to
weak  H, C  and  Mg absorption  lines, suggesting  the  spectrum of  a
stellar source.

\section{Discussion}

\subsection{Colour data and the problem of the distance}
 
Figure~\ref{fig:cmd} shows  the $Y$  vs. $(Z-Y)$  and the  $K_{\rm s}$
vs. $(J-K_{\rm  s})$ color-magnitude diagrams  (CMDs) for a  10 arcmin
radius region around the position of VVV-WIT-07.  According to the VVV
extinction  maps   \citep{2012A&A...543A..13G}  this  region   has  an
extinction  of $A_K=1.85$  mag (integrated  along the  entire line  of
sight),  corresponding  to  $E(J-K)=3.51$~mag,  assuming  the  law  of
\cite{2009ApJ...696.1407N}.   The CMDs  suggest that  VVV-WIT-07 is  a
Galactic object located in the foreground disk region.  The VVV colors
are roughly constant  over time (see Table 1 for  the observed epochs)
and consistent either  with a reddened MS star  \citep[spectral type G
  or earlier,  e.g.][]{2018arXiv180806139A} or a pre-MS  star, but the
surroundings do not  show evidence of an active  star formation region
(see Fig.~\ref{fig:image}).

The proper motion  (PM) of VVV-WIT-07 as measured by  the VVV InfraRed
Astrometric    Catalogue    \cite[VIRAC,][]{2018MNRAS.474.1826S}    is
$\mu=2.826\pm0.863$~mas\,yr$^{-1}$                     ($\mu_{\alpha},
\mu_{\delta}$=$2.469,1.374$~mas\,yr$^{-1}$).   Those   values  are  as
expected for  disk field stars,  that show an asymmetric  drift.  Even
though  the  distance  of  VVV-WIT-07   is  uncertain,  the  VIRAC  PM
significantly different from zero is  an indication that the object is
relatively nearby. While photometric data for VVV-WIT-07 are available
in     Gaia     DR2      \citep[source     ID     5974962995291907584,
][]{2018A&A...616A...1G} no PM or parallaxes measurements are reported
yet for  the object in Gaia  DR2.  The lack of  motion measurements in
Gaia may suggest  a minimum distance for VVV-WIT-07,  for instance, in
\cite{2018arXiv180503171P}  Gaia data  are used  to map  the MW  disk,
including towards the position of VVV-WIT-07,  to distances of up to r
$\sim$  7 kpc,  close  to the  Galactic center.   On  the other  hand,
VVV-WIT-07 ($G$=20.56~mag) is near the  limiting magnitude of Gaia and
an archive  search in the region  around VVV-WIT-07 shows that  only a
small  fraction  ($\lesssim  1/4$)  of sources  at  similar  magnitude
($G>$20.5 mag) present measured PMs and parallaxes in Gaia DR2.

\begin{figure}
\centering
\includegraphics[scale=1.01]{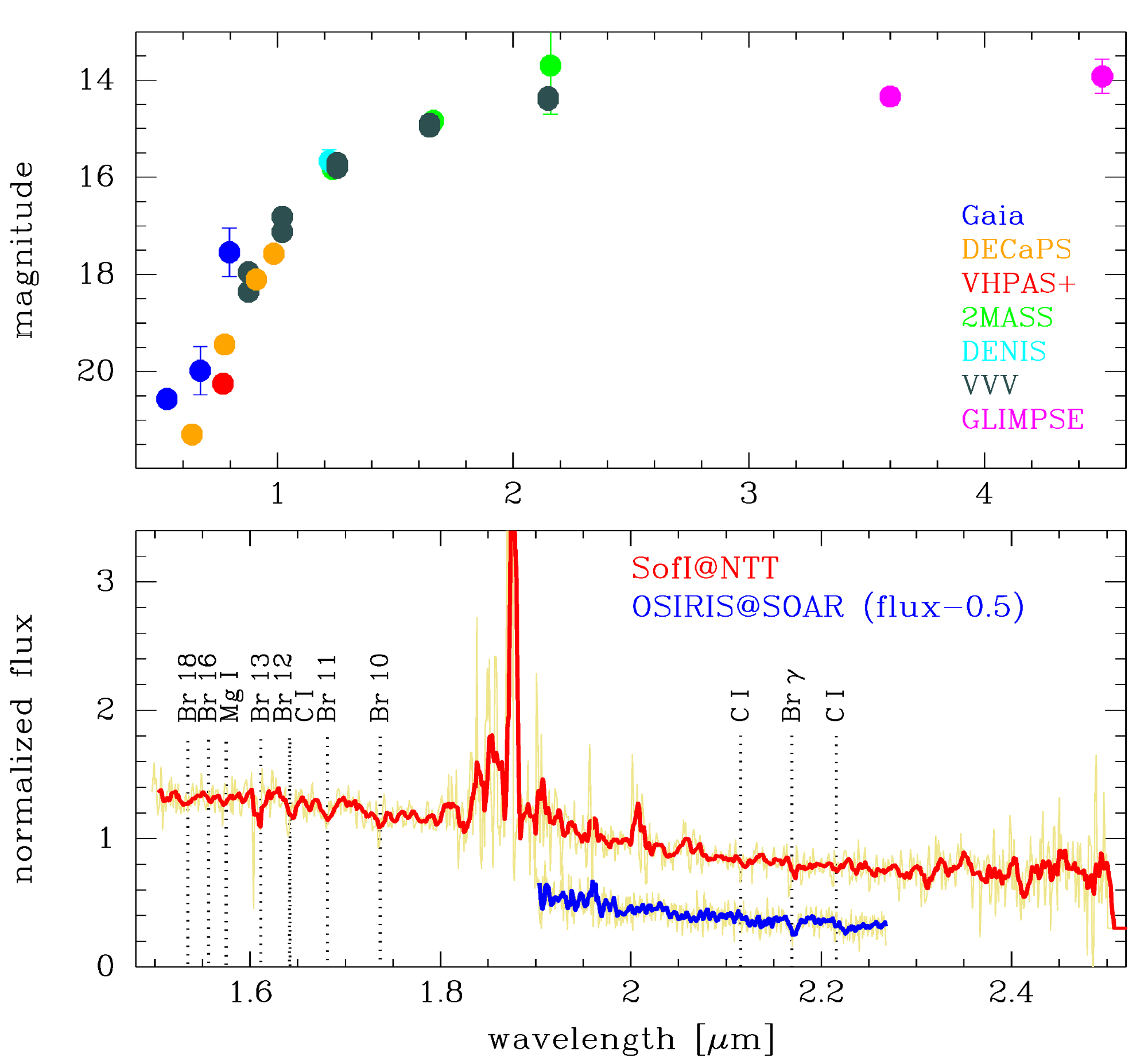}
\caption{Top: Photospectrum of VVV-WIT-007 including archive data from
  different observatories/surveys.  Data covers from optical to mid-IR
  (see Table  1).  The extinction for  a 10' region around  the target
  position  is  $A_K=1.85$~mag.   Bottom:  near-IR  spectra  taken  at
  ESO-NTT on Feb 02 2016 ($J$- to $K$-band) and at SOAR on Jun 18 2016
  ($K$-band).  The spectra  are basically flat, showing  likely weak H
  absorption  features.   Strong  telluric  lines  dominate  the  SofI
  spectrum  in  the region  $1.8~\mu  $m$~  \lesssim \lambda  \lesssim
  2.0~\mu $m.}
\label{fig:sed}
\end{figure}

For a  foreground disk object,  the total extinction as  calculated by
the VVV maps is certainly overestimated. $A_{Ks}=1.85$~mag corresponds
to $A_V>15$~mag in the optical,  turning the slope of Photospectrum of
VVV-WIT-07 presented in Fig.~\ref{fig:sed} to peak at $\lambda<0.5~\mu
m$,   translating   to   a    black-body   temperature   higher   than
$T=5400$~K. 3-D extinction  maps \citep{2014A&A...566A.120S} show that
the extinction increases linearly with the distance up to $D<8$ kpc in
the   VVV-WIT-07  direction,   making   the   interpretation  of   the
photospectrum   of  VVV-WIT-07,   presented  in   the  top   panel  of
Fig.~\ref{fig:sed}  puzzling  and  dependent   of  a  better  distance
estimation.    Moreover,   a   photospectrum  for   VVV-WIT-07   using
multi-epoch data taken  at different epochs over almost  a decade must
be seen with caution since the object is clearly variable over time.

\begin{table}
\caption[]{Archive  data  for  VVV-WIT-07.   Observations  cover  from
  optical  to  mid-IR.  The  VVV  $K_{\rm  s}$ epochs  presented  here
  correspond to the ones observed  simultaneously with the $J$ and $H$
  bands. Gaia, DECaPS and VPHAS+ observations are contemporaneous with
  the VVV/VVVX data, however, epochs for Gaia observations are not yet
  publicly available (see Section 2).}
\begin{tabular}{llccr}
\hline
Filter & Survey & $\lambda_{C}$ & Mag  & Epoch  \\
       &        & [$\mu$m]    & [mag] & [JD] \\
\hline
$G$   & Gaia DR2$^{1}$& 0.532 & $20.565\pm0.016$  & NA \\
$r$   & DECaPS$^{2}$  & 0.638 & $21.293\pm0.044$ & 2457805 \\
$BP$  & Gaia IGSL$^{3}$& 0.673 & $17.545\pm0.500$  & NA \\
$i$   & VPHAS+$^{4}$  & 0.770 & $20.25 \pm0.09 $  & 2456155 \\
$i$   & DECaPS$^{2}$  & 0.777 & $19.442\pm0.018$ & 2457805 \\
$RP$  & Gaia IGSL$^{3}$& 0.797 & $17.545\pm0.500$  & NA \\
$Z$   & VVV          & 0.878 & $18.356\pm0.040$  & 2455766  \\
$Z$   & VVV          & "     & $17.957\pm0.032$  & 2457162  \\
$z$   & DECaPS$^{2}$  & 0.911 & $18.105\pm0.007$ & 2457805 \\
$Y$   & DECaPS$^{2}$  & 0.985 & $17.574\pm0.022$ & 2457805 \\
$Y$   & VVV          & 1.021 & $17.125\pm0.021$  & 2455766  \\
$Y$   & VVV          & "     & $16.814\pm0.018$  & 2457162  \\
$J$   & 2MASS$^{5}$   & 1.240 & $15.830\pm0.010$  & 2451035 \\
$J$   & DENIS$^{6}$   & 1.221 & $15.666\pm0.230$  & 2451062 \\
$J$   & VVV          & 1.254 & $15.804\pm0.061$  & 2455326  \\
$J$   & VVV          & "     & $15.709\pm0.010$  & 2455387  \\
$J$   & VVV          & "     & $15.777\pm0.011$  & 2455411  \\
$J$   & VVVX         & "     & $15.707\pm0.011$  & 2457279  \\
$H$   & 2MASS$^{5}$   & 1.664 & $14.839\pm0.094$  & 2451035 \\
$H$   & VVV          & 1.646 & $14.953\pm0.013$  & 2455326  \\
$H$   & VVV          & "     & $14.897\pm0.014$  & 2455387  \\
$H$   & VVV          & "     & $14.961\pm0.017$  & 2455411  \\
$H$   & VVVX         & "     & $14.912\pm0.017$  & 2457279  \\
$K$   & 2MASS$^{5}$   & 2.164 & $13.702^{*}$~~~~~~~~\, & 2451035 \\
$K_s$  & VVV          & 2.149 & $14.356\pm0.022$  & 2455326  \\
$K_s$  & VVV          & "     & $14.402\pm0.025$  & 2455387  \\
$K_s$  & VVV          & "     & $14.384\pm0.021$  & 2455411  \\
$K_s$  & VVVX        & "     & $14.368\pm0.024$  & 2457279  \\
3.6   & GLIMPSE$^{7}$ & 3.545 & $14.338\pm0.175$  & 2453742 \\
4.5   & GLIMPSE$^{7}$ & 4.442 & $13.923\pm0.354$  & 2453742 \\
\hline
\end{tabular}
\small{$^{1}$\cite{2018A&A...616A...1G};
  $^{2}$\cite{2017arXiv171001309S};   $^{3}$\cite{2013yCat.1324....0S}
  $^{4}$\cite{2016yCat.2341....0D};  $^{5}$\cite{2003tmc..book.....C};
  $^{6}$\cite{2005yCat.2263....0D};  $^{7}$\cite{2009yCat.2293....0S};
  $^{*}$upper limit.}
\end{table}

Both  SofI  and  SOAR  near-IR  spectra  are  featureless,  having  no
prominent lines  in emission or absorption,  excluding the possibility
of  VVV-WIT-07  as an  emission  line  object  such as  a  cataclysmic
variable (CV) or a Nova star.  The absence of $H\alpha$ data in VPHAS+
would confirm  this interpretation.   The shallow  absorption features
are interpreted as  H, C and Mg absorption lines,  that reinforces the
hypothesis of a main sequence stellar source.

\begin{figure}
\centering
\includegraphics[scale=0.89]{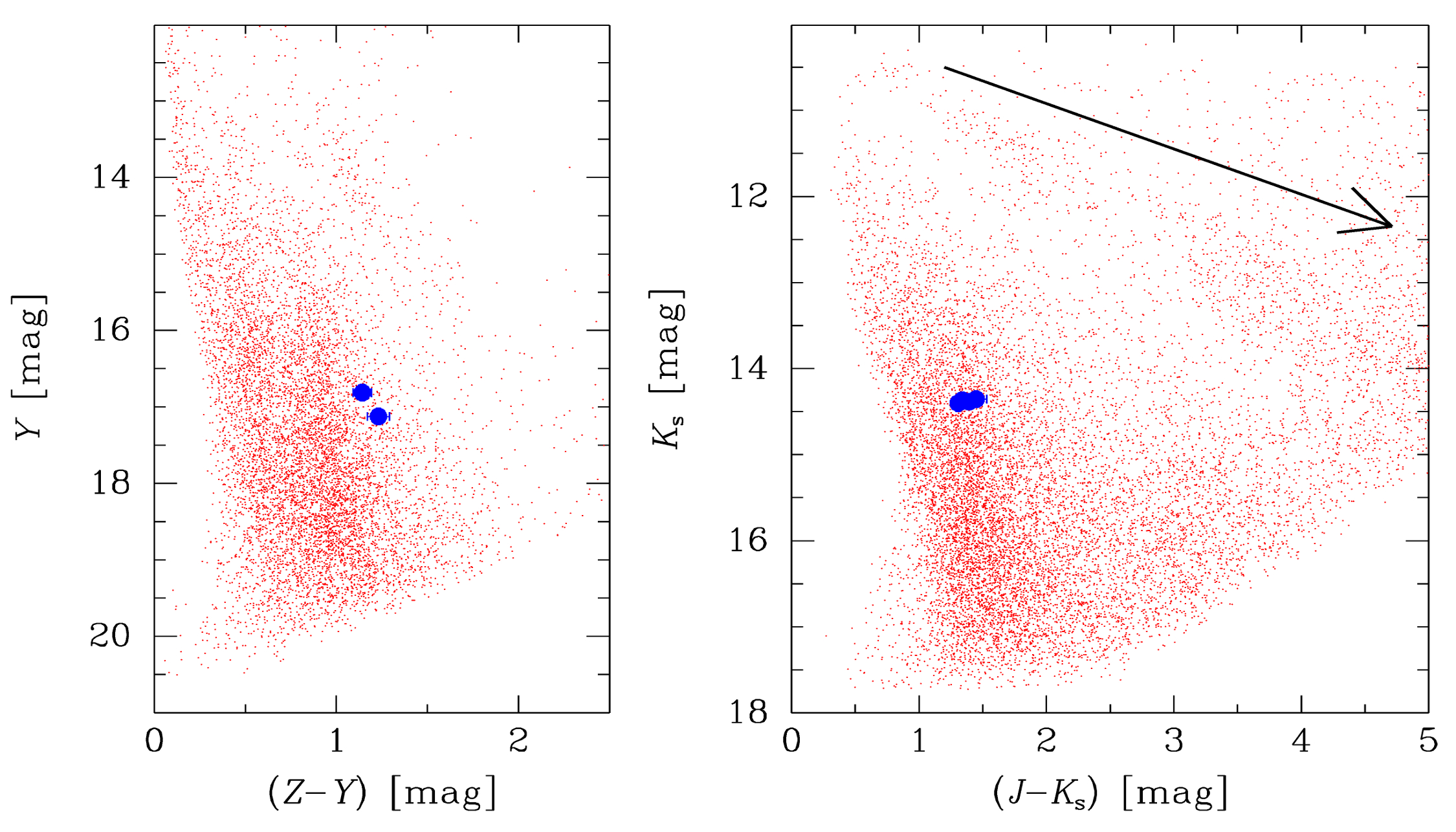}
\caption{$Y$  vs.   $(Z-Y)$ (left)  and  $K_{\rm  s}$ vs.   $(J-K_{\rm
    s})$CMDs (right) for stellar sources within 10 arcmin radii around
  the target  position.  Colors for  VVV-WIT-07 from the VVV  and VVVX
  surveys  are marked  as blue  dots:  two sets  of $ZY$  and four  of
  $JHK_{\rm s}$ colors, taken at different epochs (see Table~1).  High
  extinction limits the number of sources  in the $Y$ vs. $(Z-Y)$ CMD.
  In  the $K_{\rm  s}$ vs.   $(J-K_{\rm  s})$ CMD  a reddening  vector
  corresponding to  an extinction  of $A_{Ks}=1.85$~mag,  assuming the
  \citet{2009ApJ...696.1407N} extinction law, is also shown.}
\label{fig:cmd}
\end{figure}

We fit the  photoespectrum using the Virtual  Observatory SED Analyzer
\citep[VOSA,  ][]{2008A&A...492..277B} with  the  disclaimer that  the
photometry was collected  over nearly a decade and may  be affected by
the source's variability.  In this  analysis we masked out Gaia, 2MASS
K-band and VPHAS+  data, since 2MASS K-band magnitude is  at the upper
limit and Gaia and VPHAS+ are close to their detection limits.

We     made     use     of      Kurucz     and     BT-Settl     models
\citep{1997A&A...318..841C,2012RSPTA.370.2765A}       with       solar
metallicity, $T_{eff}$ between 3500 and 15000~K and $log\,g$ between 0
and  5,   and  assuming  $A_V$  between   $1-12$~mag  ($A_{Ks}\lesssim
1.4$~mag).   For   this  exercise  we  obtained   either  $A_v=5$  and
$T_{eff}=3,500$~K and $log\,g=5$ which  is unlikely because that would
mean a nearby  red dwarf ($d<300$~pc for a M1  dwarf). The position on
the CMDs, the  spectra, the PM and  lack of a parallax  do not support
this result. The other solution was $A_V\sim 9$~mag with a best fit in
$T_{eff}\sim 10,000$~K  and log\,$g\sim 3$,  slightly high for  an A0V
star.   If that  is the  case and  we assume  a regular  A0V star  the
distance will be $\gtrsim 3.5$~kpc.

Allowing for even higher  extinction of $A_v=15$~mag (corresponding to
$A_{Ks}\sim  1.8$~mag) leads  to $A_V  \sim~9$ mag,  with $T_{eff}\sim
9,000$~K and $log\,g\sim 2.5 - 3$.  This implies the presence a nearby
(d$\lesssim 3.5$~kpc) dust cloud with  extinction, contrary to the 3-D
extinction maps that show a  smooth, linear increment of $A_{Ks}$ with
the       distance       in       this       line       of       sight
\citep[e.g. ][]{2014A&A...566A.120S}.

\begin{figure}
\centering
\includegraphics[scale=0.9]{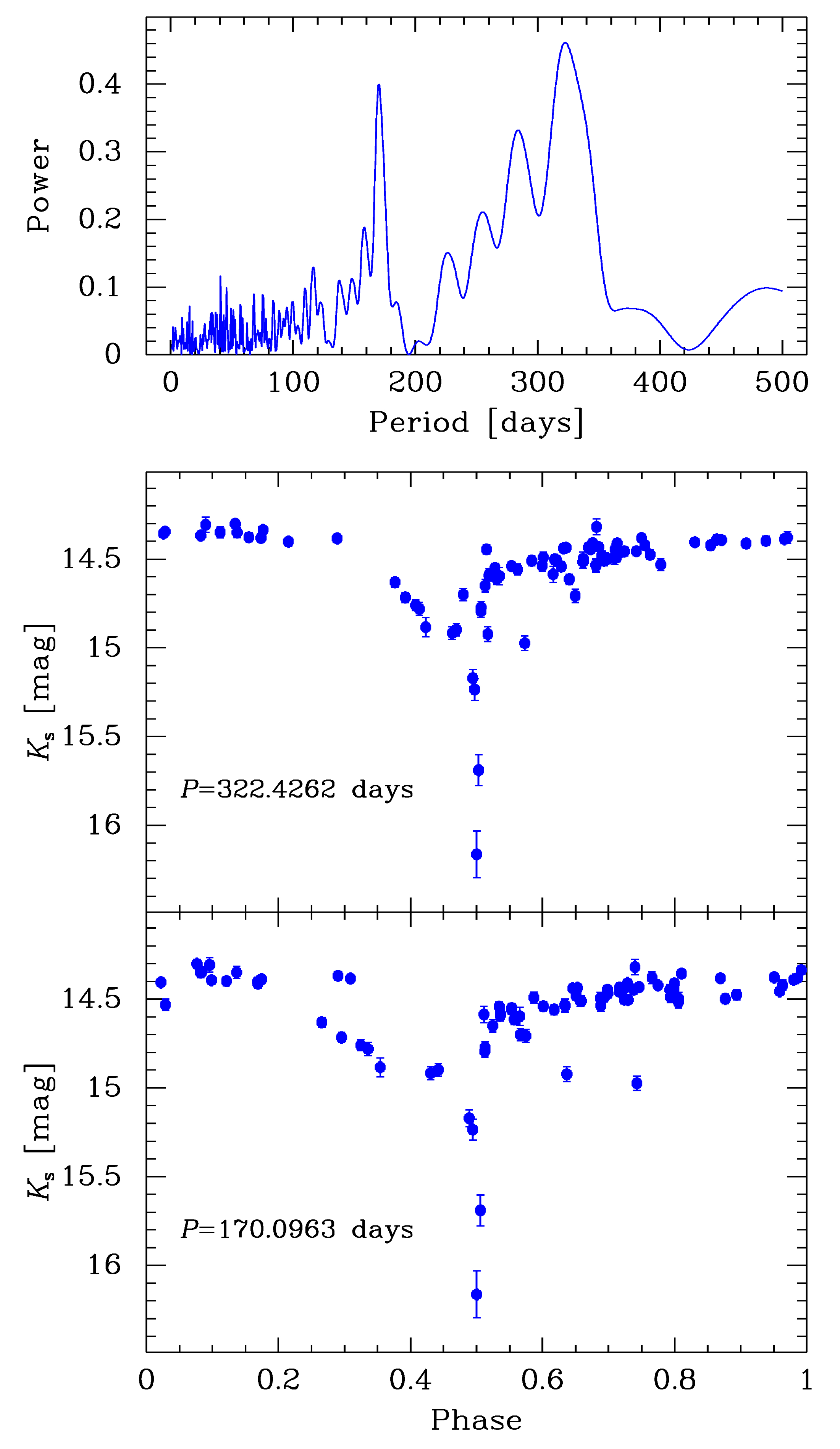}
\caption{Periodogram (top panel) and  phase-folded light curve for the
  tentative periods  of $P\sim322$~days (middle panel,  score 0.462 in
  the periodogram) and $P\sim170$~days (bottom panel, score 0.400).}
\label{fig:phase}
\end{figure}

\subsection{To be or not to be periodic?}

The light curve of VVV-WIT-07 presented in Fig.~\ref{fig:lcurve} shows
a sequence  of dips, with  a likely deep  eclipse in July  2012. While
these are certainly  recurrent, the presence of  a regular periodicity
is not  obvious.  A search  for periodicity using a  fast Lomb-Scargle
algorithm  \citep[][]{2015ApJ...812...18V}  results in  two  tentative
periods at  $P\sim322$~days (score  0.462) and  $P\sim170$~days (score
0.400).    Figure~\ref{fig:phase}  shows   the  periodogram   and  the
phase-folded  light curves  for  both  periods.   Despite  its  higher
significance, a period of $P\sim322$~days breaks the shape of the main
eclipse/dip of July  2012 by including a data point  at $K_{\rm s}\sim
14.7$~mag at the ingress of  the event ($\phi\sim0.48$), moreover data
points  seen to  spread out  near phase  $\phi\sim0.65-0.70$.  On  the
other hand, for the period of $P\sim170$~days a couple of outlier data
points  are  seen prior  ($\phi=0.3$)  and  after ($\phi\sim0.75$  and
$0.75$) the main  eclipse/dip.  Thus, we are not  able to conclusively
establish a period for VVV-WIT-07.  If the $P\sim170$~days is correct,
an  ephemeris  for  the  object  can be  calculated  to  predict  that
VVV-WIT-07  should  fall  down  to  $K_{\rm  s}\gtrsim  14.7$~mag  and
possibly become fainter than $16$~mag in $K_{\rm s}$-band, three times
in the  next year,  at around  January 21, July  10, and  December 27,
2019. In the case of  $P\sim322$~days another eclipse/dip should occur
around August 7, 2019.

\section{Possible interpretations}

The features  found in  the light curve of  VVV-WIT-07 are  similar to
those  seen in  J1407  \citep[=  V1400 Cen,  ][]{2012AJ....143...72M}.
J1407  is  consistent with  a  pre-MS  K5  dwarf  with a  ring  system
eclipsing the star.  Its light curve  has a main eclipse of $>3.3$~mag
and multiple  dimming events of  $>0.5$ mag.  During the  main eclipse
the object  fades from  $V\sim12.4$ to $V\sim15.8$,  or a  decrease of
$\sim95\%$ in the flux, compared  with $\sim80\%$ of VVV-WIT-07 in the
near-IR.   The main  eclipses  are  similar in  shape,  with a  smooth
ingress and  a shoulder in the  egress, typical for the  eclipse of an
extended  object such  as  a  disk or  ring,  which  is the  preferred
explanation in the case of J1407. If VVV-WIT-07 has the same nature of
the Mamajek object, which has a lower limit on the period of 850 days,
the  absence of  a  firm  period determination  of  VVV-WIT-07 can  be
explained by the irregular cadence  of our light curve, which presents
large windows of no coverage, especially in the last 2 years.

Another object that  is possibly similar to VVV-WIT-07  is KIC 8462852
\citep[Boyajian's star,][]{2016MNRAS.457.3988B}.   The object is  a F3
IV/V star showing  irregular and aperiodic dips, however  the dips are
shallow at  20\% of the  normalized flux, compared with  $\sim80\%$ of
VVV-WIT-07. Boyajian's star  has been followed up  since its discovery
and the  most accepted hypothesis for  its dips is the  occultation by
orbiting   material  \citep[e.g.    uneven   rings   of  dust,   dusty
  planetesimals     or     even      a     swarm     of     comet-like
  bodies][]{2016MNRAS.457.3988B,2017A&A...600A..86N}.     A   possible
period for  Boyajian's star is  still unknown, with some  recent works
suggesting  the   presence  of   a  likely  recurrence   ranging  from
$P\sim1600$~days  up  to  $P\sim12$~years  \citep{2017RNAAS...1...33B,
  2017arXiv171001081S, 2018MNRAS.473L..21B}.

Alternative scenarios for VVV-WIT-07 include a ``dipper'' T Tauri star
with clumpy  dust structures orbiting  in the inner disk  that transit
our line of sight \citep[e.g. ][]{2017ApJ...848...97R}, or even a long
period,  high-inclination  X-ray  binary.  The  deep,  narrow  eclipse
delayed with  respect to a broad  and shallower dip is  reminiscent of
the  morphology  seen  in  high-inclination  low-mass  X-ray  binaries
\citep[LMXB,
  e.g.][]{1986ApJ...308..199P,2002MNRAS.335..665B}. However, LMXBs are
restricted to orbital periods of less  than a few days while high-mass
x-ray binaries (HMXB) can be found at $P_{orb}$ up to hundreds of days
\citep[e.g.                        X1145-619                       has
  $P_{orb}=187.5\,d$,][]{1981MNRAS.195..197W}.    Moreover,  in   this
scenario optical and  IR spectra would be dominated  by the mass-donor
companion  star,  and   should  show  rotationally-broadened  hydrogen
absorption lines at epochs of no  mass ejection episodes, which is not
the seen in the spectra of VVV-WIT-07.
 
An R Coronae Borealis (R  CrB) classification could be also suggested,
but the light curve of VVV-WIT-07  does not resemble what is typically
observed in other  R CrB systems in the near-IR,  specially because of
the fast dimming episode in July  2012.  Moreover, the position in the
CMD  is  unlikely  for an  R  CrB.   In  particular,  if this  were  a
high-luminosity R CrB star undergoing unusually fast dimming episodes,
VVV-WIT-07 would be  located well beyond the Milky Way  disk, which is
inconsistent   with  the   measured  PM   and  amount   of  foreground
reddening. For instance, if VVV-WIT-07 has the same absolute magnitude
of         the          R         CrB          class         prototype
($M_K\simeq-6$~mag)\footnote{simbad.u-strasbg.fr/simbad/sim-id?Ident=R+Coronae+Borealis}
that would translate to a distance to VVV-WIT-07 of $d \gtrsim 50$~kpc
(using $m_{K_{\rm s}}=  14.35$~mag and $A_K=1.85$ mag,  see Sections 2
and 3).

\section{Conclusions}

We  have presented  the  discovery of  VVV-WIT-07,  a unique  variable
source  located   in  the  Galactic   disk,  identified  in   the  VVV
Survey.  VVV-WIT-07 is  a Galactic  object located  in the  foreground
disk. Its light-curve shows  dimming episodes, resembling the features
seen in Boyajian's star and  in Mamajek's object.  Other possibilities
also relate to the young stellar zoo or the eclipse of extended bodies
in a MS  star. At present, with  the information at hand,  none of the
proposed scenarios can  be conclusively established. In  any case, all
of these possibilities are interesting in their own right.

If this  is another Mamajek  object, it  means that these  objects are
likely  more common  than  previously  realized, as  is  shown by  the
example     of     the     discovered    of     OGLE     LMC-ECL-11893
\citep{2014ApJ...797....6S},      an     eclipsing      B9III     star
($P_{orb}$=468~days) consistent  with a dense circumstellar  dust disk
structure; and PDS 110, an eclipsing  system with likely transits by a
companion  with  a circumsecondary  disc  \citep{2017MNRAS.471..740O}.
Indeed,     near-IR     surveys     like    VVV     and     UKIDSS-GPS
\citep{2008MNRAS.391..136L}  have only  recently,  after a  deliberate
search, been found  to contain a number  of intriguing large-amplitude
YSOs  \citep{2017MNRAS.465.3011C,2017MNRAS.472.2990L}.  Thus,  surveys
like ours, apart of course from its irregular cadence, may perhaps not
have found objects  like WIT-VVV-07 more often  primarily because they
were not looking  specifically for this kind of  variability. The next
generation of  synoptic surveys  such as LSST,  WFIRST and  PLATO will
certainly be  major contributors  to this  field, yielding  many other
interesting discoveries.

\section*{Acknowledgements}

We gratefully acknowledge  the use of data from the  ESO Public Survey
program IDs 179.B-2002 and 198.B-2004  taken with the VISTA telescope,
and data products from the  Cambridge Astronomical Survey Unit (CASU).
This  publication  makes use  of  VOSA,  developed under  the  Spanish
Virtual Observatory project supported  from the Spanish MINECO through
grant  AyA2017-84089.  R.K.S.   acknowledges support  from CNPq/Brazil
through  projects 308968/2016-6  and 421687/2016-9.   Support for  the
authors is provided  by the BASAL CONICYT Center  for Astrophysics and
Associated  Technologies  (CATA)  through grant  AFB-170002,  and  the
Ministry  for   the  Economy,   Development,  and   Tourism,  Programa
Iniciativa Cient\'ifica Milenio through grant IC120009, awarded to the
Millennium Institute of Astrophysics (MAS). D.M.  acknowledges support
from FONDECYT  through project Regular \#1170121.   F.G.  acknowledges
support from  CONICYT-PCHA Doctorado  Nacional 2017-21171485  and from
Proyecto FONDECYT  REGULAR 1150345.   C.C.  acknowledges  support from
from  ICM N\'ucleo  Milenio de  Formaci\'on Planetaria,  NPF and  from
project  CONICYT  PAI/Concurso  Nacional  Insercion  en  la  Academia,
convocatoria  2015,  folio  79150049.  M.C.   gratefully  acknowledges
additional support by Germany's DAAD  and DFG agencies, in addition to
FONDECYT    grant    \#1171273    and   CONICYT/RCUK's    PCI    grant
DPI20140066.  R.A.  acknowledges  financial support  from  the  DIDULS
Regular PR17142 by Universidad de La Serena. The authors would like to
thank   Eric   Mamajek  for   the   helpful   suggestions  about   the
interpretation of VVV-WIT-07 data.







\appendix

\section{VVV-WIT-07 light curve}

Here  we present  the 85  $K_{\rm s}$-band  data-points of  VVV-WIT-07
available from VVV/VVV-X  and used to build the  light curve presented
in  Fig.   2.  The  photometric  flag  in  all measurements  is  $-1$,
corresponding    to    a    stellar    source    as    described    in
\cite{2012A&A...537A.107S}.

\FloatBarrier
\begin{table}
\centering
\begin{tabular}{cc}
\hline
MJD    & $K_{\rm s}$-band \\
(days) & (mag)           \\
\hline
\texttt{55326.3237} & \texttt{14.356\,$\pm$\,0.022} \\
\texttt{55387.2660} & \texttt{14.402\,$\pm$\,0.025} \\
\texttt{55411.1085} & \texttt{14.384\,$\pm$\,0.021} \\
\texttt{55484.0358} & \texttt{14.446\,$\pm$\,0.026} \\
\texttt{55690.3995} & \texttt{14.377\,$\pm$\,0.025} \\
\texttt{55696.3680} & \texttt{14.381\,$\pm$\,0.024} \\
\texttt{55697.3311} & \texttt{14.336\,$\pm$\,0.022} \\
\texttt{55795.0961} & \texttt{14.700\,$\pm$\,0.034} \\
\texttt{55807.0662} & \texttt{14.923\,$\pm$\,0.041} \\
\texttt{55825.1012} & \texttt{14.974\,$\pm$\,0.041} \\
\texttt{56084.1562} & \texttt{14.630\,$\pm$\,0.027} \\
\texttt{56089.2537} & \texttt{14.716\,$\pm$\,0.028} \\
\texttt{56094.2113} & \texttt{14.760\,$\pm$\,0.030} \\
\texttt{56096.0677} & \texttt{14.781\,$\pm$\,0.037} \\
\texttt{56099.1924} & \texttt{14.884\,$\pm$\,0.054} \\
\texttt{56112.2212} & \texttt{14.917\,$\pm$\,0.036} \\
\texttt{56114.1422} & \texttt{14.898\,$\pm$\,0.035} \\
\texttt{56122.1067} & \texttt{15.171\,$\pm$\,0.048} \\ 
\texttt{56123.0269} & \texttt{15.235\,$\pm$\,0.060} \\
\texttt{56124.0031} & \texttt{16.164\,$\pm$\,0.132} \\
\texttt{56124.9651} & \texttt{15.690\,$\pm$\,0.087} \\ 
\hline
\end{tabular}
\end{table}

\addtocounter{table}{-1}

\begin{table}
\centering
\begin{tabular}{cc}
\hline
MJD    & $K_{\rm s}$-band \\
(days) & (mag)          \\
\hline
\texttt{56126.1122} & \texttt{14.794\,$\pm$\,0.033} \\ 
\texttt{56126.1728} & \texttt{14.772\,$\pm$\,0.033} \\
\texttt{56128.2039} & \texttt{14.650\,$\pm$\,0.036} \\
\texttt{56130.1034} & \texttt{14.593\,$\pm$\,0.030} \\ 
\texttt{56130.2159} & \texttt{14.584\,$\pm$\,0.029} \\
\texttt{56132.9995} & \texttt{14.558\,$\pm$\,0.027} \\
\texttt{56133.0766} & \texttt{14.550\,$\pm$\,0.026} \\
\texttt{56134.0311} & \texttt{14.612\,$\pm$\,0.031} \\
\texttt{56135.1495} & \texttt{14.596\,$\pm$\,0.050} \\
\texttt{56141.1675} & \texttt{14.540\,$\pm$\,0.026} \\
\texttt{56144.0130} & \texttt{14.559\,$\pm$\,0.030} \\
\texttt{56151.0198} & \texttt{14.510\,$\pm$\,0.028} \\
\texttt{56156.0239} & \texttt{14.537\,$\pm$\,0.032} \\
\texttt{56162.1016} & \texttt{14.502\,$\pm$\,0.026} \\
\texttt{56163.0295} & \texttt{14.504\,$\pm$\,0.024} \\
\texttt{56175.9971} & \texttt{14.516\,$\pm$\,0.034} \\
\texttt{56176.0734} & \texttt{14.498\,$\pm$\,0.037} \\
\texttt{56188.0512} & \texttt{14.498\,$\pm$\,0.027} \\
\texttt{56202.0374} & \texttt{14.456\,$\pm$\,0.024} \\
\texttt{56213.9876} & \texttt{14.532\,$\pm$\,0.033} \\ 
\texttt{56478.9848} & \texttt{14.491\,$\pm$\,0.031} \\ 
\texttt{56488.9741} & \texttt{14.439\,$\pm$\,0.025} \\ 
\texttt{56490.2115} & \texttt{14.436\,$\pm$\,0.025} \\ 
\texttt{56501.0208} & \texttt{14.435\,$\pm$\,0.027} \\ 
\texttt{56502.0810} & \texttt{14.445\,$\pm$\,0.024} \\ 
\texttt{56503.1196} & \texttt{14.410\,$\pm$\,0.023} \\ 
\texttt{56505.0062} & \texttt{14.319\,$\pm$\,0.044} \\ 
\texttt{56506.0725} & \texttt{14.432\,$\pm$\,0.025} \\
\texttt{56513.9928} & \texttt{14.447\,$\pm$\,0.031} \\
\texttt{56514.0887} & \texttt{14.487\,$\pm$\,0.032} \\
\texttt{56515.0051} & \texttt{14.443\,$\pm$\,0.029} \\
\texttt{56515.1139} & \texttt{14.411\,$\pm$\,0.024} \\
\texttt{56527.0019} & \texttt{14.382\,$\pm$\,0.022} \\
\texttt{56531.1395} & \texttt{14.475\,$\pm$\,0.028} \\
\texttt{56553.0009} & \texttt{14.405\,$\pm$\,0.022} \\
\texttt{56566.0279} & \texttt{14.393\,$\pm$\,0.023} \\
\texttt{56578.0104} & \texttt{14.412\,$\pm$\,0.025} \\
\texttt{56806.2868} & \texttt{14.586\,$\pm$\,0.046} \\
\texttt{56810.2445} & \texttt{14.542\,$\pm$\,0.027} \\
\texttt{56814.1102} & \texttt{14.614\,$\pm$\,0.029} \\
\texttt{56817.0782} & \texttt{14.707\,$\pm$\,0.037} \\
\texttt{56827.1137} & \texttt{14.533\,$\pm$\,0.034} \\
\texttt{56827.2208} & \texttt{14.537\,$\pm$\,0.037} \\
\texttt{56830.0180} & \texttt{14.480\,$\pm$\,0.025} \\
\texttt{56831.1165} & \texttt{14.509\,$\pm$\,0.030} \\
\texttt{56836.2591} & \texttt{14.496\,$\pm$\,0.035} \\
\texttt{56837.2257} & \texttt{14.490\,$\pm$\,0.026} \\
\texttt{56838.0742} & \texttt{14.447\,$\pm$\,0.025} \\
\texttt{56838.1785} & \texttt{14.468\,$\pm$\,0.025} \\
\texttt{56841.1300} & \texttt{14.457\,$\pm$\,0.025} \\
\texttt{56851.1179} & \texttt{14.422\,$\pm$\,0.023} \\
\texttt{56883.1825} & \texttt{14.422\,$\pm$\,0.029} \\
\texttt{56886.1011} & \texttt{14.390\,$\pm$\,0.026} \\
\texttt{56910.0692} & \texttt{14.398\,$\pm$\,0.026} \\
\texttt{56919.1077} & \texttt{14.388\,$\pm$\,0.026} \\
\texttt{57279.0012} & \texttt{14.368\,$\pm$\,0.024} \\
\texttt{57584.1264} & \texttt{14.346\,$\pm$\,0.020} \\
\texttt{57926.3172} & \texttt{14.307\,$\pm$\,0.042} \\
\texttt{57933.3156} & \texttt{14.349\,$\pm$\,0.033} \\
\texttt{58210.3311} & \texttt{14.379\,$\pm$\,0.033} \\
\texttt{58263.2300} & \texttt{14.301\,$\pm$\,0.025} \\
\texttt{58210.3311} & \texttt{14.379\,$\pm$\,0.033} \\
\texttt{58263.2300} & \texttt{14.301\,$\pm$\,0.025} \\
\texttt{58264.1774} & \texttt{14.350\,$\pm$\,0.029} \\
\hline
\end{tabular}
\end{table}

\bsp	
\label{lastpage}
\end{document}